\documentclass[sigconf]{acmart}
\usepackage[T1]{fontenc}
\usepackage[utf8]{inputenc}
\usepackage{lmodern}
\usepackage[most]{tcolorbox}
\usepackage{tabularx, booktabs, stfloats}
\usepackage{comment}

\newcommand{\commentEx}[1]{\textcolor{CadetBlue}{\textbf{\textit{%
  \fontsize{9pt}{10pt}\selectfont#1}}}}

\AtBeginDocument{%
  \providecommand\BibTeX{{%
    \normalfont B\kern-0.5em{\scshape i\kern-0.25em b}\kern-0.8em\TeX}}}

\setcopyright{acmcopyright}
\copyrightyear{2024}
\acmYear{2024}
\acmDOI{XXXXXXX.XXXXXXX}

\acmPrice{15.00}
\acmISBN{978-1-4503-XXXX-X/18/06}

\begin{document}

\title{DocGen: Generating Detailed Parameter Docstrings in Python}

\author{Vatsal Venkatkrishna}
\authornotemark[1]
\email{vatsalvenkatkrishna@gmail.com}
\affiliation{%
  \institution{Australian National University}
  \country{Australia}
}
\author{Durga Shree Nagabushanam}
\email{durgashree.n15@gmail.com}
\affiliation{%
  \institution{Australian National University}
  \country{Australia}
}
\authornote{Both authors contributed equally to this research.}

\author{Emmanuel Iko-Ojo Simon}
\email{ammanuel.Simon@anu.edu.au}
\affiliation{%
  \institution{Australian National University}
  \country{Australia}
}
\author{Melina Vidoni}
\email{melina.vidoni@anu.edu.au}
\affiliation{%
  \institution{Australian National University}
  \country{Australia}
  }
\renewcommand{\shortauthors}{Venkatkrishna and Nagabushanam, et al.}

\begin{abstract}
Documentation debt hinders the effective utilization of open-source software. Although code summarization tools have been helpful for developers, most would prefer a detailed account of each parameter in a function rather than a high-level summary. However, generating such a summary is too intricate for a single generative model to produce reliably due to the lack of high-quality training data. Thus, we propose a multi-step approach that combines multiple task-specific models, each adept at producing a specific section of a docstring. The combination of these models ensures the inclusion of each section in the final docstring. We compared the results from our approach with existing generative models using both automatic metrics and a human-centred evaluation with 17 participating developers, which proves the superiority of our approach over existing methods.
\end{abstract}





\maketitle

\section{Introduction}

%\FHF{In the introduction, you should convince the reader that your study is important and tackles something needed and significant, so your article is worth being accepted and read. You should discuss the following:}

%\FHF{What is the problem and why is this problem important (include why Python only)?}

Documentation is a vital aspect of software development, but it is often neglected by developers because of its tedious and time-consuming nature \cite{shmerlin2015document}. As a result, documentation becomes incomplete, outdated or non-existent over time \cite{vidoni2022understanding}, leading to various problems such as increased maintenance costs, reduced software quality, and lower user satisfaction \cite{aghajani2019software}. 
This is particularly concerning for scientists and mathematicians who develop software tailored to their specific goals. Since they are rarely trained in software engineering and coding best practices, they tend to leave poor documentation for ``scientific software'' \cite{segal2007some, pinto2018scientists}. There is a tremendous increase in the inclusion of software applications for scientific research \cite{nguyen2010survey}, simultaneously requiring extensive documentation to be useful and reproducible \cite{pawlik2012documentation}. In the context of open-source software, method-level documentation (e.g., Python’s docstrings) is crucial, as it enables the effective utilization of the software for further development and maintenance \cite{pressman2005software}.  Python is one of the most widely used programming languages in computational science due to its versatility, simplicity and efficiency \cite{oliphant2007python, ayer2014scientists}.

Over the years, source code summarization \cite{haiduc2010use, mcburney2014automatic, mcburney2015automatic} has been studied extensively, aiming to produce a high-level summary of a given code snippet. Several pre-trained language models specifically developed for code understanding tasks \cite{codebert, codet5, unixcoder} have shown promising results on code summarization and a significant improvement over classical text-only summarization algorithms \cite{niu2023empirical}. However, there have been few studies that analyse these algorithms' capability of generating a ``detailed docstring'' --- one that describes the entire function holistically, with all the technical details of its parameters, which is essential for understanding and reusing code \cite{sridhara2011generating}. This is significantly more challenging than code summarization since it requires an in-depth understanding of each individual parameter's role in the function, as well as their interactions with other variables. Moreover, human-written documentation is often deficient \cite{lethbridge2003software}, making high-quality training data scarce and limiting the performance of generative models. Therefore, complex documentation needs such as thoroughly explained parameters in dynamically typed languages remain an open challenge. 

In this work, we present an approach to generate documentation catering to the specific needs of scientific software. We based our approach on the paper by \citet{Vidoni2023taxonomy}, which developed a taxonomy of parameter-level documentation directives\footnote{A documentation directive is a natural-language statement to inform developers of constraints and guidelines related to the correct usage of a section of code; in our case, a function's parameter}. Using this taxonomy as a baseline, we devised a composite framework with multiple modules, each addressing a different documentation directive. We compared our results to those obtained from using a single transformer-based model through automatic scoring methods like BLEU \cite{papineni-etal-2002-bleu} and METEOR \cite{banerjee-lavie-2005-meteor} and further validated our approach through a survey with 17 participating developers.

The rest of this paper is organized as follows. We present prior work that inspires and closely relates to our own (\S\ref{sec: related-works}). We then formalize our research questions (\S\ref{subsec: rqs}) and describe the data-gathering steps along with an inclusion-exclusion criterion (\S\ref{subsec: data_gathering}). We then detail the issues observed in training data (\S\ref{subsec: issues}) and subsequent preprocessing steps (\S\ref{subsec: data-proc}). We then describe our approach of independently generating each directive and the design choices we made (\S\ref{subsec: model_prep}). Finally, we discuss the results we obtained (\S\ref{sec: results-discussion}) and its implications (\S\ref{sec: implications}), along with threats to validity (\S\ref{sec: threats}) and a conclusion (\S\ref{sec: conclusion}).

%We identified the issues in human-written documentation which can prevent the pre-trained models from generating all the technical details for parameter documentation and hence the need for the multi-model approach. This approach will come in handy in enhancing the quality of parameter docstring generation. 

%\FHF{I would add a short discussion in the introduciton and related work about why you did not use LLMs, and also the reasons of though LLMs are so powerful, they (might) not be able to solve the problem. You can search for some works and see if there is any discussions about the need for PLMs (i.e., pre-trained language models such as codebert) and/or the drawbacks of LLMs to support your discussion. }

\section{Related Works}
\label{sec: related-works}
%\FHF{Think what is the purpose of this section. Divide in subsections (with just \textbf{bold} text). There is a section on comment generation, a section on importance of Python and related works to yours on Python only. Probably a section on LLM. At the end, include one paragraph stating \textit{explicitly} how your work is different from the existing ones. }

\textbf{Deep learning for source code summarization.}
Code summarization is the task of generating a natural language description, typically a single line, for a code snippet describing its purpose and functionality. Early approaches treated source code as plain text and employed machine translation algorithms to generate summaries \cite{iyer-etal-2016-summarizing, woo2018cbam}. However, programming languages have complex syntax and rich structure, which make them fundamentally different from natural language. Several studies have employed methods to exploit the structural and semantic information of source code through Abstract Syntax Trees (ASTs) \cite{alon2019code2seq, zheng2019codeattention, leclair2019neural, hu2018deep}, Control Flow Graphs (CFGs) \cite{beyene2021source} and Data Flow Graphs (DFGs) \cite{guo2021graphcodebert} for code summarization. In recent years, models pre-trained jointly on code and natural language have been very successful, with CodeT5 \cite{codet5}, ProphetNet-Code \cite{qi2021prophetnetx}, CoTexT \cite{phan2021cotext}, SPT-Code \cite{niu2022sptcode} and UniXcoder \cite{unixcoder} being some of the best performing models for code summarization \cite{niu2023empirical}.

\textbf{Docstring generation.}
\citet{cui2022codeexp} introduced the task of ``code explanation generation,'' which aims to generate an informative and detailed summary of the code. They provided an annotated dataset as well as a curriculum learning pipeline, achieving promising results. However, their codebase was not reproducible at the time of performing our experiments (May 2023), with the dataset they collected being inaccessible and irreproducible due to missing files and a lack of documentation. Moreover, their approach does not guarantee the generation of important docstring directives that we address through our approach. \citet{clement2020pymt5} developed an approach inspired by the T5 architecture \cite{raffel2023exploring} and achieved promising results on generating good quality docstrings, but similarly lack details on reproducibility.

\textbf{Method-level documentation.} 
\citet{barone2017parallel} developed a large corpus of Python code-docstring pairs by mining open-source repositories on GitHub and trained on NMT models to obtain baseline results. \citet{sulir2017generating} approached method documentation by obtaining examples for parameters, returns and object states before and after method execution and converted the essence into natural language.
Among the specific directives we cover in this paper, to the best of our knowledge, only datatype prediction has been extensively studied in prior work.
\citet{luo2018recognizing} proposed a method, training a classifier for type prediction. \citet{pradel2020typewriter} proposed a similar neural classifier with an additional framework to ensure compatibility of predicted types. \citet{peng22hityper} integrated static prediction with deep learning through a dependency graph generated for an entire given repository. However, all these methods rely on both existing documentation and source code for producing predicting datatypes, which introduces a cyclic dependency on documentation. \citet{mir2022type4py} developed a hierarchical neural network for type inference through similarity learning. 

\textbf{Taxonomies for documentation.} 
The increasing number of taxonomies has contributed to the maturing of the field of Software Engineering \cite{usman2017taxonomy}. Taxonomies on software documentation contain a detailed account of directives for developers to include in their repositories. Embedding such a taxonomy into docstring generation tools could improve the quality of documentation generated in terms of technical details and consistency. \citet{monperrus2012should} performed an empirical study of projects in Java to develop a taxonomy of API directives. They introduce the ``method call'' sub-directive, which specifies the directives of a method to be included in API documentation. \citet{Vidoni2023taxonomy} developed a taxonomy for R documentation, which specifically classifies the directives for method-level documentation.

\textbf{Novelty of our approach.} We propose a multi-step fine-tuning approach to produce detailed docstrings documenting every parameter in a function while ensuring the inclusion of four directives -- a description of the parameter's overall purpose, its datatype, its default value, and its acceptance of a ``None'' value. Since we used different models for each of these directives, we were able to use only those models that were best suited to each directive's need. Our choice of these directives is informed by a filtered version of the taxonomy introduced by \citet{Vidoni2023taxonomy}, covering the most important and challenging directives presented. However, the aforementioned taxonomy was designed specifically for scientific software. Since different fields of software would benefit from a different set of directives being documented, we limited the scope of our study to scientific software to avoid presuming the nature of developers' needs at large. 
% We explored docstring generation by narrowing down our scope to scientific software, and Python in particular; the reason to do this was to keep the scope manageable while focusing on the current most-used programming language\footnote{According to IEEE Spectrum Ranking: \url{https://spectrum.ieee.org/top-programming-languages-2022}}. 
% Moreover, as this is an exploratory study, we focused only on the documentation of parameters due to their relevance for scientific software \cite{hannay2009scientists}.
We conducted experiments to verify the novelty of our approach through automated scoring methods like BLEU \cite{papineni-etal-2002-bleu} and METEOR \cite{banerjee-lavie-2005-meteor} as well as through a survey of 17 experienced developers who compared our results with existing fine-tuning approaches.
%The training data is a collection of human-written documentation, prone to numerous inconsistencies as discussed in \ref{add}. Hence, a single model, on training, cannot exercise to generate all the attributes of a parameter including the description, type, default status, and acceptance of none values.

\section{Methodology}
\label{sec: methodology}

\subsection{Directives to Document}
\label{subsec: taxonomy}
A taxonomy of documentation directives could be very helpful in guiding best practices. However, to our knowledge, there exists no taxonomy that describes the details to be documented for Python functions. However, such a taxonomy does exist for R \cite{Vidoni2023taxonomy}. Considering similar aspects of both Python and R, like dynamic typing and seamlessly blending object-oriented programming and functional programming, we decided to use this taxonomy as a guide for our models. To maintain simplicity in our work, we selected four directives from the taxonomy based on the availability of high-quality training data, namely:
\begin{itemize}
\item PD: a description of the general purpose and functionality of a parameter
\item PV: the default value of a parameter
\item PT: the default datatype of a parameter
\item PN: Whether or not a parameter can take on a ``None'' value (the original taxonomy tracks both ``Na'' and ``null'' in R, which is equivalent to Python's ``None'').
\end{itemize}
Other directives, like exceptions that could be raised and the data formats of primitive and non-primitive datatypes, are equally crucial to a detailed docstring but were rarely included in our datasets. Hence, we chose to use the directives where data is either available or could be easily extracted.

\subsection{Research Questions (RQs)}
\label{subsec: rqs}
%-description -relevancy -methodology
\textbf{RQ1. Can each directive be produced using a separate model best suited to the task's complexities?}
Parameters are an important part of modern programming, especially in dynamically typed languages like Python, allowing users to reuse a function multiple times with varying inputs. Hence, it is crucial to document them systematically to allow for maintenance and reusability. We investigate the best approaches to learning the intricacies of generating the directives described in \S\ref{subsec: taxonomy}.
% Parameters are reportedly essential for data science software since they are used to configure algorithms and models implemented as functions, which frequently provide the results as a return. Therefore, it is fundamental to develop comprehensive documentation. This paper proposes leveraging the selected taxonomy \cite{Vidoni2023taxonomy} to build the different directives composing each parameter's docstring. To do this, rather than expecting one model to produce all the directives, we aim to experiment with a multi-step strategy with a combination of summarization and classification models.

\textbf{RQ2. Is a combination of task-specific models a more effective approach for docstring generation compared to a single model for all directives?} 
For a single generative model to reliably produce all required docstring directives, the corresponding training data must contain a majority of instances with these directives present. Unfortunately, such code-docstring pairs are lacking in quantity and hard to enforce without manual annotation. Thus, we propose a multi-step fine-tuning method that guarantees the generation of these directives and compare it with existing single-model approaches.
% Principally, generative models are pre-trained either on Natural Language (NL) or on code and structures. Docstring's conventions are not sufficiently exercised by developers during documentation \cite{rani2021comments}. Therefore, fine-tuning these models for a structured generation task using the existing repositories developed across the globe in numerous formats may not suffice for the inclusion of all the derivatives of the taxonomy. Further, the inclusion of various directives of a parameter is not just a generative task. This paper assesses a multi-step approach to discern between the various constituents of a docstring, setting to path to fine-tune models effectively. 
    
\textbf{RQ3. Do developers prefer the docstrings generated by our proposed multi-step approach over a single model's docstring?}
Developers are the primary benefactors of the software documentation. Thus, their opinions on the docstrings generated by our approach are highly valued. We conducted a survey with a Likert-based scoring strategy to assess their opinion on the docstrings from our multi-step approach and contrast them with the docstrings generated by a single generative model.

\subsection{Data Gathering}
\label{subsec: data_gathering}

\textbf{Inclusion-Exclusion Criteria}
We defined inclusion and exclusion criteria (IEC) to determine suitable repositories for our study; they are summarised in Table \ref{tab: incl-excl}.

\begin{table}[h]
\centering
\caption{Inclusion and exclusion criteria (IEC) for selecting relevant software repositories.}
\label{tab: incl-excl}
\footnotesize
\begin{tabularx}{\linewidth}{c X}
\toprule
\textbf{Code} & \textbf{Inclusion Criteria} \\
\midrule
I1 & Repositories with documentation written in English \\
I2 & Repositories comprising at least 75\% Python code \\
I3 & Repositories with licenses among Apache license 2.0, Creative Commons license family, BSD Zero-Clause license, GNU General Public License family, MIT license\\
I4 & Scientific software \\
\midrule
\textbf{Code} & \textbf{Exclusion Criteria} \\
\midrule
E1 & Repositories with at least one commit after 1st Feb 2023 \\
E2 & Functions with no parameters \\
E3 & Functions with no return variables \\
E4 & Jupyter notebooks \\
E5 & Forked repositories \\
\bottomrule
\end{tabularx}
\end{table}

We considered repositories with English as their primary language [I1] to avoid loss of semantics during translation. To limit the number of functions with dependencies on files written in other programming languages, we ensured that the repositories under consideration constituted at least 75\% of all code written in Python  [I2]. To avoid breaches in copyright regulation, we only mined repositories with selected licenses [I3]. We only included repositories that have one of (\texttt{data-science, machine-learning, deep-learning, statistics, and science}) listed as topics to restrict our study to scientific software \cite{zhang2022code} [I4].

We considered inactivity as an exclusion criterion [E1] to ensure the possibility of submitting a pull request to repositories lacking any documentation for extended validation on the docstrings generated if required. We also excluded the functions that do not accept parameters or return any value [E2, E3] due to their prominent role in a docstring \cite{Vidoni2023taxonomy}. We further excluded Jupyter notebooks due to differences in their documentation standards \cite{grotov2022large} [E4] and forked repositories to avoid duplication [E5]. We did not place any restrictions on the accompanying documentation in the IEC to allow for a testing set.

\textbf{A dataset of scientific software.}
We used the dataset introduced by \citet{boa}, which contains 1558 repositories that use a diverse set of data science libraries. However, we still needed to check these repositories against our inclusion-exclusion criteria. We employed a two-stage approach for this check. First, we extracted the metadata pertaining to our IEC for all 1558 repositories using \texttt{libraries.io}\footnote{\url{https://libraries.io/api}} because of its high rate limit (60 requests per minute). However, the API was unable to retrieve the data for many existing repositories. Therefore, we used GitHub's REST API\footnote{\url{https://docs.github.com/en/rest?apiVersion=2022-11-28}}, which has a lower rate limit (1000 requests per hour), to retrieve the repositories that failed in \texttt{libraries.io}. After filtering, we were left with 873 repositories that use libraries related to data science and fit our IEC. 

However, these 873 repositories had a very small number of repositories listing topics like \texttt{science} and \texttt{statistics}. Therefore, we manually mined repositories to represent other forms of scientific software. We applied our IEC through GitHub's Advanced Search Engine to search for repositories. This search retrieved $2422$ repositories, but on closer inspection, we found a substantial number of false positives. Thus, two authors manually analyzed $200$ of the retrieved repositories with the most stars and classified them on the basis of compliance with IEC. The authors individually provided a verdict of ``one'' indicating compliance and a ``zero'' otherwise for each repository. We achieved a Cohen's Kappa score of $0.779$, denoting high inter-rater reliability between the authors to include $109$ of the $200$ repositories into the final dataset, resulting in a total of $982$ repositories.

%\FHF{This section is confusing. You first say we collected 2600 repos. Then say we keep 109 out of 200 repos. Then say we use BOA. What did you do at the end? From 2600 repos you collected, how many did you include? Is BOA in addition?}
\textbf{Extracting functions and docstrings.}
We extracted all the functions from our dataset of $982$ repositories using \texttt{function-parser}\footnote{\url{https://nathancooper.io/function_parser/}}, resulting in $111$k  functions with their corresponding docstrings. However, not all of these docstrings contained documentation corresponding to each parameter. Thus, we filtered these to include docstrings which contain at least one of the following tokens -- \texttt{``:param'', ``:arguments'', ``:args'', ``:parameters'', ``param:'', ``arguments:'', ``args:'', ``parameters:''}, since they are commonly observed in docstring conventions. Although this approach may have led to a few false negatives, it ensured the absence of false positives, thus maintaining the quality of our dataset. This filtering step leaves us with $41,462$ functions, which we hereby refer to as the \textit{Raw} dataset.

% \textbf{Data Quality Checks}
% The selected repositories provided 111k functions with docstrings; however, upon close inspection, not all of them had parameters (either on the function declaration or in the docstring). Therefore, we filtered this sample to retain only the code-comment pairs that included parameters and that were parseable as an AST, a tree representation of the syntactic nature of source code. The reason for the latter is that the pre-trained models are generally fine-tuned on function code using an AST-based approach for description generation. This yielded 15.6k functions, which we used as our dataset.  

\subsection{Issues with Existing Docstrings}
\label{subsec: issues}

We randomly sampled 50 human-generated docstrings from our \textit{Raw} dataset to identify the best data preprocessing strategies forward. We made the following observations on the issues with human-generated documentation, which supported the need for our multi-step approach to generate parameter docstrings: 
\begin{figure}[h!]
    \centering
    \includegraphics[width=0.478\textwidth]{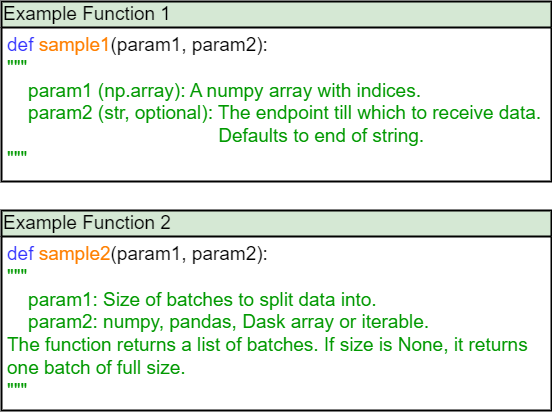}
    \caption{Two sample docstrings, highlighting much of the issues we found in the \textit{Raw} dataset. The docstrings differ in format and in the amount of information documented, a problem that becomes harder to deal with an increasing number of samples.}
    \label{fig: problems}
\end{figure}
\begin{itemize}
    \item The amount of information documented for each parameter is not constant -- In \textit{Example Function 1} in Figure \ref{fig: problems}, \textit{param2} has an ``optional'' attribute, indicating the presence of a default value, whereas the default value status of \textit{param1} is not addressed. 
    \item Some technical details are mentioned as part of the general description -- In \textit{Example Function 2} of Figure \ref{fig: problems}, the ``None'' status of \textit{param1} and its corresponding action is mentioned as part of the short summary of the docstring and is left out when parameter descriptions alone are used for training. 
    \item The position in which the datatype of a parameter is mentioned in the docstring may not be consistent -- In Figure \ref{fig: problems}, the datatype of parameters is mentioned in parenthesis (\textit{Example Function 1}), mentioned in the description or is not mentioned at all (\textit{Example Function 2}).
\end{itemize}
A single generative model might struggle with these issues because it has to decide what details to include or exclude without any guidance. A multi-step approach solves this problem by using specific directives that ensure the inclusion of certain information. Producing a single directive is easier than producing a whole docstring because of persistent documentation debt. Moreover, each directive has its own intricacies, which a model trained for that specific directive could learn better.

\subsection{Data Preprocessing}
\label{subsec: data-proc}
\begin{figure}[h!]
    \centering
    \includegraphics[width=0.478\textwidth]{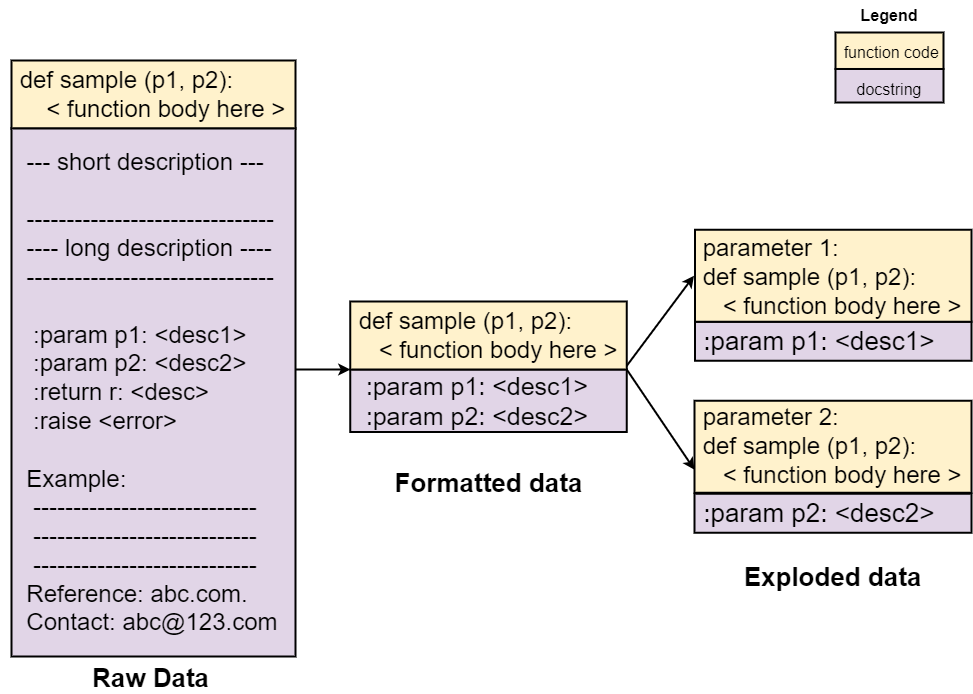}
    \caption{An example of the preprocessing of \textit{Raw} data to obtain the \textit{Formatted} and \textit{Exploded} datasets on a single instance. We discarded contacts, external links, short and long descriptions, return descriptions and potential errors raised from the \textit{Raw} data and unified all parameter docstrings under a single format to obtain the \textit{Formatted} data. We then split the instance into its constituent parameters and prepend the phrase ``parameter N'' before each code snippet.}
    \label{fig: preprocessing}
\end{figure}

The following subsections detail our preprocessing approach and are summarized visually in Figure \ref{fig: preprocessing}

\textbf{Standardizing the data format.}
Docstrings can follow several writing conventions (e.g., ReST, Google, Numpydoc-style), and a mix of different styles in the data could negatively affect our models. Prior work suggests that the content documented takes precedence over the format it is presented in \cite{rani2021comments}. Therefore, to ensure formatting consistency in the training data, we cleaned up the \textit{Raw} dataset by discarding newline tokens, extra spaces and external links/email addresses. We also removed examples enclosed by some docstrings along with long descriptions to ensure focus on parameters. 

Further, we extracted the names, descriptions and datatypes of all parameters present in the docstring using \texttt{docstring-parser}\footnote{\url{https://pypi.org/project/docstring-parser/}}. This resulted in a dataset constituting parameter names and their corresponding docstrings. 

%committing grammatical or spelling errors.
\textbf{Discarding partially documented functions.}
Upon closer inspection of our data, we found that some functions were only partially documented. For example, consider the following function and its docstring \commentEx{def precook(s, n=4, out=False) ":param s: string, sentence to be converted into n-grams. :param n: int, number of n-grams for which representation is calculated."}. The description for the third parameter, ``out'', is not included. Therefore, we developed an algorithm to extract the parameters from function headers and compared it to the list of parameters extracted by \texttt{docstring-parser}. The results were ranked as follows:

\begin{itemize}
\item \textbf{Rank one:} The number of parameters extracted by \texttt{docstring-parser} is equal to the number of parameters extracted by our parameter extractor. 

\item \textbf{Rank two:} The number of parameters extracted by \texttt{docstring-parser} is greater than zero and lesser than the number of parameters extracted by our parameter extractor.

\item \textbf{Rank three:} The number of parameters extracted by \texttt{docstring-parser} is zero, and the number of parameters extracted by our parameter extractor is greater than zero.
\end{itemize}
We only considered the functions ranked one for further pre-processing as these followed the docstring writing conventions \cite{browning2014docstring}, promising a better quality of data. We denote this dataset as the \textit{Formatted} dataset, which we used as a benchmark for the best possible performance through a single model.

\textbf{``Exploding'' the data.}
Our proposed multi-step approach generates the documentation for one parameter at a time, thereby requiring the training data for our models to be re-formatted accordingly. We split the \textit{Formatted} dataset into $N$ data points for each existing function, where $N$ is the number of parameters in the corresponding function. To avoid different outputs for the same code snippet, we prepended the phrase ``parameter \textit{N}: '' to the code string in each entry as shown in Figure \ref{fig: preprocessing} to indicate the parameter being documented in the docstring. This decomposition had a dual effect of simplifying the task as well as increasing the amount of training data. However, as highlighted in \S\ref{subsec: issues}, these descriptions sometimes contained auxiliary information which, while seemingly harmless, could confuse the model into learning a pattern of these inclusions that doesn't actually exist. Thus, we restricted each entry in the dataset to only the first sentence. We hereby refer to this dataset as the \textit{Exploded} dataset because it was obtained using the \texttt{pandas.DataFrame.explode} module\footnote{\url{https://pandas.pydata.org/docs/reference/api/pandas.DataFrame.explode.html}}. Both the Formatted and Exploded datasets were split into training and validation sets, with 20\% data in the training set. All fine-tuning is performed on the respective training sets, and results are on the respective validation sets.

\subsection{Model Preparation}
\label{subsec: model_prep}
\begin{figure*}[h]
    \centering
    \includegraphics[width=7in]{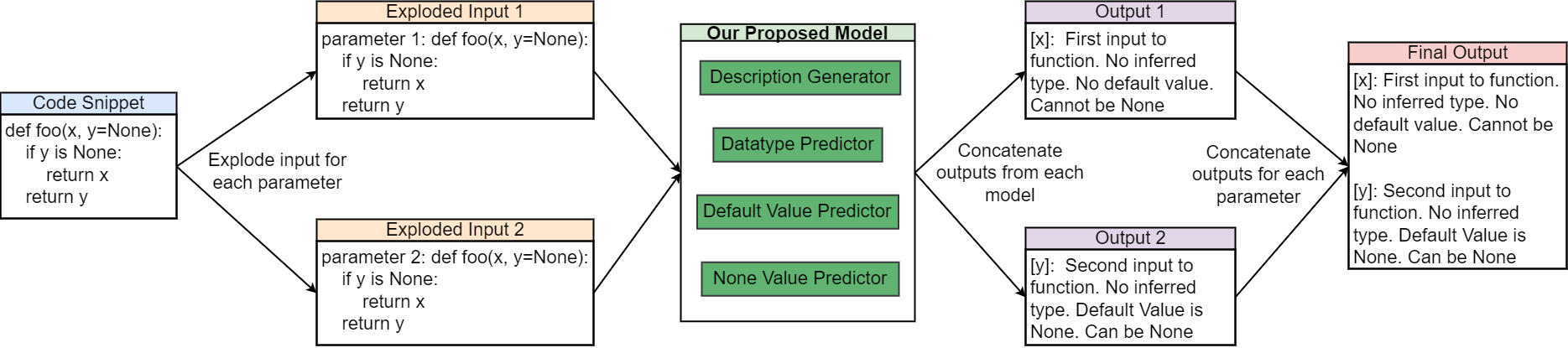}
    \caption{The pipeline for our proposed multi-step approach. Each input code snippet is replicated \textit{N} times, where \textit{N} is the number of parameters. Each replication is tagged with its corresponding parameter number. Each of these replications was then passed through the multi-model pipeline, comprising a description generator, datatype predictor, default value predictor, and None acceptance classifier. We concatenated the outputs from these models to produce our final output.}
    \label{fig:testing-pipeline}
\end{figure*}
Pre-trained language models (PLMs) are deep learning models with millions, often billions, of parameters that can be fine-tuned to perform a variety of downstream tasks \cite{wang2022pre}. PLMs have also been applied to a number of code-understanding tasks to great success \cite{niu2022deep} like code summarization, which generates a single-line summary of a code snippet highlighting its purpose \cite{codebert, codet5, unixcoder}. This enables us to fine-tune PLMs to perform our task of docstring generation for parameters. To our knowledge, very few studies have investigated the performance of these models in generating a detailed docstring, which is a much harder task than code summarization, and none have proposed a similar multi-step approach. We excluded Large Language Models (LLMs) from this analysis to limit the scope of our study to a multi-step approach and not digress into a debate about the benefits and demerits of using LLMs. We intend to perform an in-depth comparison of our approach with popular LLMs like GPT-4 \cite{OpenAI2023GPT4TR} in a future study. Below, we describe our approach to producing each directive in detail.

%We primarily focused on documenting the parameters of a function because they are more in number and have a greater chance of being documented. We referred to the taxonomy proposed by \citet{Vidoni2023taxonomy} while choosing different directives of our output docstring. 
\textbf{Parameter Descriptions (PD).}
The task of generating parameter descriptions is similar to code summarization but requires an understanding of each parameter's role in the function. We experimented with three PLMs that are trained jointly on natural and programming languages:

\begin{itemize}
\item CodeBERT: \citet{codebert} were the first to present a bimodal architecture, trained jointly on both natural and programming languages. It does not consist of a Transformer decoder, which must be trained from scratch for generative downstream tasks like code summarization. We included this model in our study as a baseline to compare with other models.

\item CodeT5: \citet{codet5} extended the T5 model architecture \cite{t5} to code-understanding and generation tasks. We included this model in our study because of its proven excellence in code summarization \cite{niu2023empirical}

\item UniXcoder: \citet{unixcoder} used mask attention matrices with prefix adapters to improve the representation of code using ASTs and code comments. We included this model in our study because it has achieved state-of-the-art in several code-related tasks \cite{niu2023empirical}.
\end{itemize}

We used the publicly available checkpoints on HuggingFace for all three models in our experiments, trained for 10 epochs with all hyperparameters at their default values. We then trained these models on the \textit{Exploded} dataset with early stopping based on the BLEU score \cite{papineni-etal-2002-bleu}. 

For a baseline comparison, we trained these models on the \textit{Formatted} dataset, which is not instructed to generate any specific directives. This helped us assess the ability of current generative models to produce detailed docstrings and contrast their outputs with those from our multi-step finetuning approach.

\textbf{Parameter Datatypes (PT).}
In the absence of good documentation, predicting each input parameter's datatype is a challenging task, requiring an in-depth understanding of their roles and interactions with other variables. Most prior work used existing documentation \cite{pradel2020typewriter, luo2018recognizing} and other files in the same repository \cite{peng22hityper} to generate datatypes.  Since we wanted to generate this documentation itself, we chose algorithms that use only the source code for predicting datatypes. To the best of our knowledge, there exists only one such algorithm -- Type4Py \cite{mir2022type4py}. It provided an API for type prediction, whose response contained data types of all parameters and variables used in the function. We extracted the predicted data types and mapped them with the corresponding parameters.

\textbf{Parameter Default Values (PV).}
Identifying default values is a relatively simple task, especially in Python, where defaults are mostly declared in the function header itself.
Thus, we implemented a simple AST-based algorithm to parse the source code and infer defaults. We used Python's inbuilt \texttt{ast} module\footnote{\url{https://docs.python.org/3/library/ast.html}} to parse the input function and return an \texttt{ast.arg} object. If there are any default values for the parameters listed in \texttt{args}, the \texttt{defaults} attribute holds them. For keyword-only arguments (listed in the \texttt{kwonlyargs} attribute), the corresponding defaults are listed in the \texttt{kw\_defaults} attribute. We extracted these default values. 

\textbf{Parameter None Acceptance (PN).}
We used a binary classification objective to classify a parameter's acceptance of the ``None'' value. We modified our \textit{Exploded} dataset to include the classification label. If the full description (beyond one line) of the parameter contained the ``None'' token, a label of $1$ was assigned, and $0$ otherwise. This resulted in an unusually large number of negative instances, which we balanced out by sampling an equal number of positive and negative instances. We fine-tuned UniXcoder and CodeBERT on the resulting dataset for 10 epochs with a learning rate of 1e-5. Although CodeT5 is capable of performing classification-based tasks \cite{codet5}, we excluded it from this part of our study to maintain a comparison between encoder-only models. We evaluated the model with our validation set using the micro F1 score.

\textbf{Final Output.} The output from each of the modules described above was concatenated to obtain the docstring of a single parameter. The docstrings of each parameter were further concatenated to obtain the final output of our multi-step approach.
% Why did we pick these models in specific? 
% - all the models you selected and why
% - experiments to decide which ones to keep
% - pipeline
% - hyperparams
% - rounds of training
% \subsection{PR Checks}

% \subsection{Replication Package}
% \label{subsec: rep-pack}
% The replication package accompanying this manuscript can be found at \url{https://tinyurl.com/2p8wpe49}. This package contains our dataset files, the scripts we used for fine-tuning and scoring these models and the files necessary to reproduce the survey. Further instructions on using the package are documented in the accompanying README file. 
% %\FHF{Are you open sourcing your dataset and/or models? If not, why? If yes, make sure the links are ALL anonymous, as the paper would be rejected immediately if authors reveal their names in any means. }

\section{Results \& Discussion}\label{sec: results-discussion}
This section describes the results obtained from our experiments and also discusses lessons learned. We present our analysis for each RQ as follows:

\subsection{RQ1. Multiple models tailored to generate specific directives can produce a detailed docstring.}
\label{sec: results-rq1}

The training data is a collection of human-written documentation which, as expected, is prone to numerous inconsistencies, as discussed in \S\ref{subsec: issues}. Therefore, we hypothesized that a single model couldn't produce docstrings with certain directives reliably generated. Our proposed approach leveraged the strengths of multiple models of varying complexity that generated their respective directives and, hence, a detailed docstring. This section focuses on evaluating the outputs from each module of our approach from fine-tuning on the \textit{Exploded} dataset with a multi-step approach. 

\textbf{Parameter Descriptions (PD).}
For evaluating parameter descriptions, we used BLEU \cite{papineni-etal-2002-bleu} and METEOR \cite{banerjee-lavie-2005-meteor}, which are commonly employed to evaluate summarization-oriented tasks. They use different measures to compute the overlap between the generated and target texts. To apply these metrics, we used HuggingFace Evaluate\footnote{\url{https://huggingface.co/docs/evaluate}}. These scores are reported in Table \ref{tab: desc-results} under the ``Exploded'' subscript.

\begin{table}[h!]
    \centering
    \caption{Results for generating parameter descriptions. Subscripts indicate the dataset each model was trained on. The best result in each category is highlighted in bold, and the overall best model is underlined.}
    \begin{tabular}{lcc}
        \toprule
         \textbf{Model Name} & \textbf{BLEU-4} & \textbf{METEOR}  \\\toprule
         CodeBERT\textsubscript{ Formatted} & 0.212 & 0.476 \\
         CodeT5\textsubscript{ Formatted} & \textbf{0.281} & \textbf{0.482} \\
         UniXcoder\textsubscript{ Formatted} & 0.224 & 0.462\\\midrule
         CodeBERT\textsubscript{ Exploded} & \underline{\textbf{0.310}} & \underline{\textbf{0.506}} \\
         CodeT5\textsubscript{ Exploded} & 0.278 & 0.485 \\
         UniXcoder\textsubscript{ Exploded} & 0.254 & 0.465 \\\bottomrule
    \end{tabular}
    \label{tab: desc-results}
\end{table}

\begin{figure}[h!]
    \centering
    \includegraphics[width=0.47\textwidth]{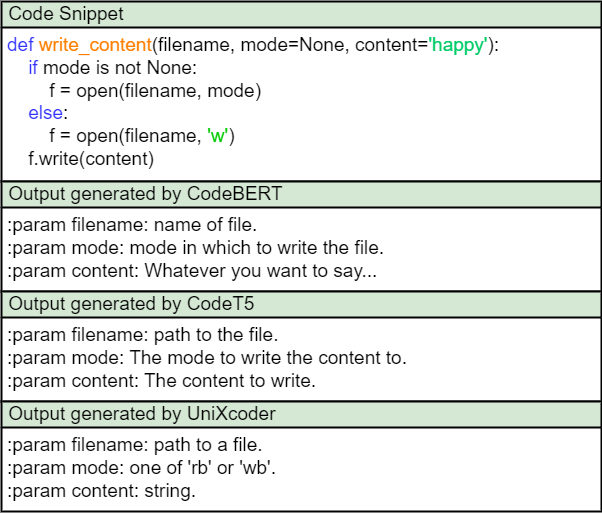}
    \caption{An example of the parameter descriptions generated from our multi-step approach using three different models.}
    \label{fig:rq1-output}
\end{figure}

We observed that CodeBERT significantly outperforms the remaining models, followed by CodeT5. This is a surprising result since CodeT5 has a decoder that is pre-trained on code-understanding tasks, while CodeBERT's decoder is fine-tuned from scratch, suggesting CodeT5's pre-training does not help in generating parameter-wise descriptions as it does in code summarization. Figure \ref{fig:rq1-output} shows a sample of the outputs generated by these models is in. The outputs of CodeBERT and CodeT5 were comparable, and the informal tone of CodeBERT's output for the \texttt{content} parameter was noteworthy. UniXcoder generates relatively bland outputs. We also observed that it points to wrong choices for the \texttt{mode} parameter in the description.

\textbf{Parameter None Acceptance (PN).}
We used the micro F1-score, a commonly used metric for binary classification \cite{sokolova2009systematic}. UniXcoder achieves a score of 0.704, followed by CodeBERT with 0.669. This behaviour is expected, since UniXcoder has outperformed CodeBERT on There is further scope for improvement, which could be achieved using a manually curated dataset instead of the synthetic one used in our approach. This will be discussed in Section \ref{rq1-fw}.

\textbf{Parameter Default Values (PV) and Datatypes (PT).}
We could not use any automatic metrics for evaluation to evaluate both parameter default values and datatypes due to the absence of a labelled ground truth. Thus, we resorted to a human-centred evaluation strategy. We applied a sample size calculation of 95\% confidence and 5\% error to determine a representative set for evaluation, sampling 369 out of 9112 instances in our validation set. A subset of 369 was then randomly sampled and inspected simultaneously by two authors independently. Both authors inferred the datatypes and default values of the parameters indicated in the source code by themselves. If their verdict matched the model's, a positive score was given. We used Cohen's Kappa coefficient to measure the level of agreement between the reviewers. Although the approach was neither comprehensive nor extensive, it has a negligible risk of researcher bias. In future works, more in-depth studies must be conducted on generating and evaluating default values and data types. 

The default values achieved a high accuracy of 93.7\%, with a Cohen's Kappa of 0.903. To understand why the accuracy falls short of 100\%, consider the following case: \commentEx{def sample3(x, y=None): if y is None: y=5}. In this case, the default value of y is actually $5$ and not ``None''. The presence of a few such samples, which are not standard Python coding practices, led to the error.

The datatype prediction achieved a significantly worse result, with an accuracy of 20\% and a Cohen's Kappa of 0.828. This was expected because predicting default values is a significantly simpler task than predicting the data types of a variable, especially in Python, which is dynamically typed. These results indicate the need for more annotated datasets for datatype prediction.

\textbf{Lessons Learned}
The performance of PLMs for specific natural language processing tasks depends on the quality of data they are fine-tuned on \cite{meng2022generating}. Documentation debt in scientific software repositories hinders us from developing all-rounder models which can include technical details along with descriptions. To establish an approach that can tackle the existing debt and still produce the required output, we split the documentation task into smaller exclusive tasks. We observed that producing individual directives was not necessarily a generative task and varied significantly in complexity. PNs were modelled as a binary classification task with a synthetic dataset, PVs performed best with a simple rule-based method, and PTs used Type4Py, which is a hierarchical neural network architecture that predicts type clusters to produce a wide variety of datatypes.

\tcbset{%
    theo/.style={%
        enhanced,
        breakable,
        sharp corners,
        toprule=0pt, rightrule=0pt, bottomrule=0pt, leftrule=1mm,
        colback=#1!5, colframe=#1!80!black, coltitle=#1!80!black, 
        detach title,
        overlay unbroken and first ={
            \node[rotate=90, minimum width=1cm, anchor=south, font=\bfseries] 
               at (frame.west) {\tcbtitle};
        }
    }
}

\newtcbtheorem[auto counter]{mytheo}{RQ}
{theo=SeaGreen}{th}

\begin{mytheo}{}{}
Models fine-tuned on code structures to produce descriptions for parameter docstrings were not consistent in the number of technical details included. Therefore, we combined the outputs from task-specific models to ensure the inclusion of the different technical directives of a parameter along with the description. 
\end{mytheo}

\textbf{Future Works}
\label{rq1-fw}
%-evaluation model - evaluation for types
The approach to evaluating parameter datatypes was neither comprehensive nor extensive, as is expected in the absence of a well-labelled dataset. The performance of multiple models in the multi-step approach was evaluated individually. A metric to assess the approach as a whole could be of interest for further improvement.

\subsection{RQ2. A combination of task-specific models is more effective than using a single model.}
We compared our approach of using multiple models to a single-model approach for parameter docstring generation. The outputs generated by a single generative model fine-tuned on the \textit{Formatted} dataset were compared to those obtained from our multi-step approach. For the single-model approach, we fine-tuned CodeBERT, CodeT5 and UniXcoder to generate the entire ground-truth docstring, as opposed to the parameter-wise input in our multi-step approach. The BLEU and METEOR scores of these approaches are shown in Table \ref{tab: desc-results}. 
We saw a significant improvement in CodeBERT's performance, almost 46\% in BLEU-4 scores. UniXcoder also saw an improvement, although not as impressive, and CodeT5 had a very minor change in performance, with a minor decrement in its BLEU-4 score. However, these results are not sufficient to suggest the superiority of our approach since it does not account for the presence of the other modules used in our approach (i.e., PV, PN and PT). Moreover, the datasets used for fine-tuning these models are also different, with the \textit{Exploded} dataset having fewer tokens than the \textit{Formatted} dataset, and hence a lower scope for error.

\begin{figure}[h!]
    \centering
    \includegraphics[width=0.478\textwidth]{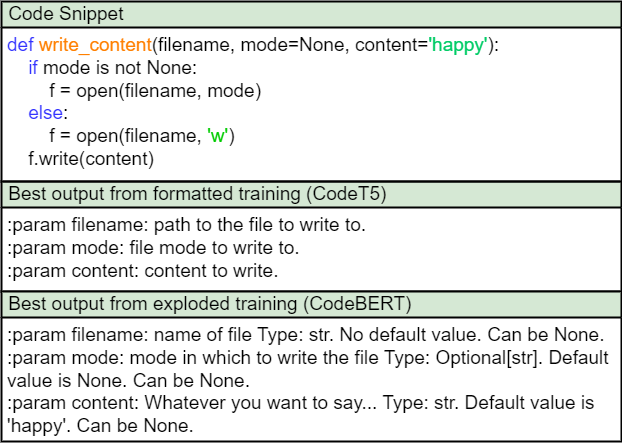}
    \caption{Comparison of outputs generated by the two approaches on the same example. The best model in both cases is showcased.}
    \label{fig:rq2-output}
\end{figure}

Figure \ref{fig:rq2-output} showcases the best output from both approaches on the same code input. We observed that the best multi-step model tended to be more descriptive and informative than a single model. However, the outputs for PT and PN are not ideal because \commentEx{filename} and \commentEx{content} cannot be ``None'' without throwing an exception, and the type for \commentEx{content} should also be \commentEx{``Optional[str]''}. Nonetheless, we argue that the multi-step output is better overall since the single-model approach only \textit{mentions} the strings ``None'' and ``Default'' 3.2\% and 1.9\% of the time, respectively. This makes the multi-step approach more capable and reliable in generating detailed docstrings, with a future scope for improvement in its individual parts.

\textbf{Lessons Learned}
The multi-step fine-tuning allows the model to relate each parameter individually with its corresponding code snippet, allowing effective learning of the respective directive. This significantly improved the quality of directives generated for each parameter. Documentation debt severely inhibits the ability of a single generative model to reliably include these directives. Additionally, our multi-step approach is more capable than a single generative model because each module is adept at handling a specific directive, allowing them to learn task-specific knowledge. 

\begin{mytheo}{}{}
Detailed docstring generation is ill-modelled as a fully generative task. Owing to an overwhelming amount of documentation debt, the quality of training data is not good enough to enable generative models to learn to include all crucial documentation directives. The difference in the underlying nature of tasks required to generate each directive of the docstring adds to the complexity. Therefore, a combination of multiple models adept at generating each directive is a more reliable approach to generating detailed docstrings. 
\end{mytheo}

\textbf{Future Works}
The taxonomy proposed by \citet{Vidoni2023taxonomy} addresses several other directives of docstrings, like exception raising and data formats, which we do not cover in this work. Further, We need to understand the nature of the tasks each of these directives requires to expand our multi-step model and incorporate suitable steps for each directive. In the future, LLMs should also be explored for their in specific docstring generation tasks. Ensuring that LLMs are actually generating these docstrings rather than retrieving them is crucial to their utility and is hard to achieve due to the large volumes of data they are trained on. 
We can also experiment with engineering prompts for LLMs to check for the possibility of achieving higher levels of completeness. 

\subsection{RQ3. Developers prefer the docstrings from our multi-step approach over those from a single model}
We perform a human-centred evaluation of our multi-step approach to accurately evaluate the performance of our approach and verify its superiority over a single model. To this end, we designed an assessment that received an ethical exemption to examine the preference of practitioners of software engineering in parameter docstrings. These practitioners were 17 graduate students in software engineering and majorly using Python in their software development projects. They were recruited through a call sent out to members of a multi-university collaboration effort (names redacted to maintain anonymity). The survey is available as part of our replication package.

We randomly sampled 10 code snippets and their corresponding docstrings generated by three different models: 1) a single model (CodeT5) generating the complete documentation, 2) CodeBERT generating description for the multi-step approach, and 3) CodeT5 generating description for the multi-step approach. We expected each surveyor to give a score between one to six on the Likert scale (i.e., from Poor to Excellent); prior research has demonstrated that six-point Likert scales reduce the uncertainty around the neutral middle point \cite{chyung2017evidence}. Our scales assessed the following criteria:
\begin{itemize}
    \item \textbf{Perceived completeness:} The presence of all the expected components of a docstring.
    \item \textbf{Understandability of descriptions:} The ease with which the parameter descriptions can be understood.
    \item \textbf{Grammatical correctness:} The correctness of sentence formations and vocabulary.
    \item \textbf{Technical nature of the content:} The usage of appropriate technical vocabulary.
    \item \textbf{Docstring understandability:} The ease with which the entire docstring can be interpreted.
\end{itemize}
\begin{table}[h!]
    \centering
    \caption{Results obtained from the survey for different criteria of evaluation.}
    \footnotesize
    \begin{tabularx}{\linewidth}{p{1.25in} p{0.5in} p{0.5in} p{0.45in}}
        \toprule
         \textbf{Criteria} & \textbf{Single model (CodeT5)} & \textbf{Multi-step (CodeBERT)} & \textbf{Multi-step (CodeT5)} \\
         \toprule
         Perceived completeness & 3.310 & 4.059 & 4.562 \\
         Understandability of descriptions & 3.558 & 4.022 & 4.457 \\
         Grammatical correctness & 4.479 & 4.324 & 4.524 \\
         Technical nature of the content & 3.673 & 4.258 & 4.544 \\
         Docstring understandability & 3.566 & 4.148 & 4.424 \\
         \bottomrule
    \end{tabularx}
    \label{tab: survey-results}
\end{table}

For the parameter descriptions (PD) in our multi-step approach, we used the outputs from CodeBERT and CodeT5, which were the two best-performing models. The other three directives are produced using the methods d in \S\ref{subsec: model_prep}. Therefore, the difference in scores between the two multi-step outputs should be interpreted as the difference in parameter descriptions generated by the models. However, the difference in scores between the single-model and multi-step approaches represents a difference in the docstrings as a whole.

From Table \ref{tab: survey-results}, we observed that developers found the outputs from the multi-step approaches more complete than those generated by a single-model CodeT5 model. Further, we observed a huge difference in scores for \textit{Understandability of Descriptions} between single-model CodeT5 generating the whole documentation and CodeT5 generating only parameter descriptions in the multi-step approach, which further indicates the overwhelming nature of docstring generation compared to parameter-wise description generation. Additionally, the inclusion of directives, like PT, PN and PV, improved the understandability. The score for \textit{Docstring Understandability} shows variation between single-model and multi-step approaches. There is no significant difference between the scores obtained in the two experiments of the multi-step approach. 

The scores for grammatical correctness for all three approaches were comparable, which could be attributed to the models being pre-trained on syntax and semantics of language. The score for the technical nature of the content also shows improvement in the multi-step approach. Overall, developers scored the multi-step approaches higher than the best single-model approach, validating our hypothesis. 

CodeT5, as a description generator in the multi-step approach, scores highest in all the cases of evaluation criteria designed for the survey. This is in direct contrast to the scores using automatic metrics in Table \ref{tab: desc-results}. This indicates that CodeT5 is more akin to developers' expectations despite CodeBERT having a higher semantic overlap with current documentation standards. Moreover, it is an indication for more human-in-the-loop evaluation metrics to be used for documentation-related tasks to ensure that language models learn to cater to the needs of developers.

\textbf{Lessons Learned.}
The results from the survey suggested that the recruited practitioners prefer our multi-step approach due to the inclusion of several docstring directives, improving their completeness. It also contributed to making the docstrings more understandable because the tasks of generating individual directives were simpler than generating an entire docstring. The technical nature of the content is also significantly greater in the multi-step approaches.
\begin{mytheo}{}{}
Practitioners prefer the output from the multi-step approach with all the technical details over output from a single model with just a description. There is sound acceptance of our approach in terms of completeness and the technical nature of the content.
\end{mytheo}

\textbf{Future Works.}
We can expand the scope of our survey by including developers from different domains of scientific software. Likewise, a Mechanical Turk approach could be used in the future, both as an assessment measure (as is done in this study) and as a data-enhancement technique (i.e., where participants improve the gold data, adding missing directives and mitigating visible cases of documentation debt).

\section{Implications}
\label{sec: implications}
\textbf{For Researchers. } 
This study provides exploratory insights into docstring documentation for parameters and serves as a baseline study for the inclusion of several technical directives into docstring generation. From discussing the contribution of documentation debt in incomplete documentation to proposing approaches to tackle the issues in training data, we have set a precedent to develop automatic docstring generation tools that mitigate future documentation debt. We add functionality and utility to the established task of code summarization. 

Within software engineering, our approach can be further extended to generating directives of returns, exceptions, and other key parts of scientific software documentation. A multi-step approach like ours offers great flexibility and versatility, which can be extended into such directives and to domains beyond software engineering for related tasks.

% We also pave the way for novel assessments on documentation completeness. In studies where a single generative model is used for documentation generation, its performance can be assessed only on the basis of the final output obtained. On the contrary, in an approach similar to ours, the conceptual validity of execution in each step of the multi-step approach can be evaluated against a clearly defined documentation directive. 

Our approach also encourages the development of taxonomies for various software use cases to enable taxonomy-guided multi-step approaches like ours and also serve as a representative gold standard for improved evaluation methods. Additionally, investigating the multilingual ability of multi-step models across programming languages could be a future study.

\textbf{For Developers and Data Scientists.}
The analysis by \citet{rani2021developers} describes the need for documentation support required by developers, especially in including relevant information corresponding to crucial directives with automated tools. Generating documentation strongly relies on the training data and, thus, on existing documentation. If developers produce incomplete documentation \cite{aghajani2019software}, the trained models will perform similarly. 

Therefore, our multi-step approach is an important step towards ensuring docstring completeness and mitigating documentation debt. This is especially relevant for dynamically typed languages, which make understanding code challenging--especially if a single variable can take on multiple datatypes and type-checks have not been coded into specific functions. Our approach is lucrative for developers and scientists looking for effective and extensive documentation generation tools. 

\section{Threats to Validity}\label{sec: threats}
We considered several validity threats that arise from the challenges associated with mining software repositories, data preprocessing and multi-step docstring generation. 

\subsection{Internal Validity Threats}
Internal validity threats arise from factors that can alter the outcome \cite{feldt2010validity}. Selection bias while mining scientific software repositories is a potential threat as repositories are mined from GitHub, a corpus containing repositories of all domains and configurations \cite{kalliamvakou2014promises}. We mitigated this threat by defining extensive inclusion and exclusion criteria. 

Researcher bias during manual annotation is another possible threat while checking for compliance of the retrieved repositories with the inclusion and exclusion criteria, especially for criteria that cannot be automated. To address this bias, two annotators validated the repositories, achieving a respectable 0.779 Cohen's Kappa score \cite{mchugh2012interrater}. Human bias during preprocessing is another potential risk to reproducibility, which we mitigated by extracting relevant directives from multiple docstring formats, thus maintaining a common format

\subsection{External Validity Threats}
External validity threats are related to the generalizability of results \cite{feldt2010validity}. 
In this study, we have considered only Python repositories in the domain of scientific software. Hence, the results can be generalized within the domain. However, we are uncertain about the applicability of the approach and results to non-scientific software repositories or scientific software repositories written in languages other than Python.

\subsection{Construct Validity Threats}
Human errors and bias could result in inaccuracies in the survey results. We address this threat by handpicking surveyors for the evaluation process, ensuring their qualifications and experience in Python for scientific software.

\section{Conclusion} 
\label{sec: conclusion}
In this exploratory study, we aimed to understand the implications of documentation debt in generating detailed docstrings for parameters. We suggest that documenting several docstring directives cannot be achieved by a single generative model. We devise a multi-step approach that uses multiple modules to generate their respective directives, thus enabling each module to learn task-specific information. We conducted a survey of practitioners to rate the documentation generated by our approach and from a single generative model to validate our hypothesis.

There are many potential aspects of the study that would benefit from future research. Firstly, one could expand the task-specific modules to include many other directives of a docstring, like error raising and data formats. One could also study the use of LLMs for generating detailed documentation. Moreover, an evaluation criterion for the whole docstring, adding to directive-specific evaluation, would be of interest.

%\section*{Acknowledgement}

%\FHF{Check ACM template and see if this is a separate section or as a footnote. Add this (and ask Dr. Mel): This research is supprted by a grant from GRANT NAME. But, don't inlcude the grant name, as everything should be anonymous.}

\bibliographystyle{ACM-Reference-Format}
\balance
\bibliography{references}


\begin{thebibliography}{60}

%%% the Web address.

\ifx \showCODEN    \undefined \def \showCODEN     #1{\unskip}     \fi
\ifx \showDOI      \undefined \def \showDOI       #1{#1}\fi
\ifx \showISBNx    \undefined \def \showISBNx     #1{\unskip}     \fi
\ifx \showISBNxiii \undefined \def \showISBNxiii  #1{\unskip}     \fi
\ifx \showISSN     \undefined \def \showISSN      #1{\unskip}     \fi
\ifx \showLCCN     \undefined \def \showLCCN      #1{\unskip}     \fi
\ifx \shownote     \undefined \def \shownote      #1{#1}          \fi
\ifx \showarticletitle \undefined \def \showarticletitle #1{#1}   \fi
\ifx \showURL      \undefined \def \showURL       {\relax}        \fi
% invisible to TeX
\providecommand\bibfield[2]{#2}
\providecommand\bibinfo[2]{#2}
\providecommand\natexlab[1]{#1}
\providecommand\showeprint[2][]{arXiv:#2}

\bibitem[Aghajani et~al\mbox{.}(2019)]%
        {aghajani2019software}
\bibfield{author}{\bibinfo{person}{Emad Aghajani}, \bibinfo{person}{Csaba Nagy}, \bibinfo{person}{Olga~Lucero Vega-M{\'a}rquez}, \bibinfo{person}{Mario Linares-V{\'a}squez}, \bibinfo{person}{Laura Moreno}, \bibinfo{person}{Gabriele Bavota}, {and} \bibinfo{person}{Michele Lanza}.} \bibinfo{year}{2019}\natexlab{}.
\newblock \showarticletitle{Software documentation issues unveiled}. In \bibinfo{booktitle}{\emph{2019 IEEE/ACM 41st International Conference on Software Engineering (ICSE)}}. IEEE, \bibinfo{pages}{1199--1210}.
\newblock


\bibitem[Alon et~al\mbox{.}(2019)]%
        {alon2019code2seq}
\bibfield{author}{\bibinfo{person}{Uri Alon}, \bibinfo{person}{Shaked Brody}, \bibinfo{person}{Omer Levy}, {and} \bibinfo{person}{Eran Yahav}.} \bibinfo{year}{2019}\natexlab{}.
\newblock \bibinfo{title}{code2seq: Generating Sequences from Structured Representations of Code}.
\newblock
\newblock
\showeprint[arxiv]{1808.01400}~[cs.LG]


\bibitem[Ayer et~al\mbox{.}(2014)]%
        {ayer2014scientists}
\bibfield{author}{\bibinfo{person}{Vidya~M Ayer}, \bibinfo{person}{Sheila Miguez}, {and} \bibinfo{person}{Brian~H Toby}.} \bibinfo{year}{2014}\natexlab{}.
\newblock \showarticletitle{Why scientists should learn to program in Python}.
\newblock \bibinfo{journal}{\emph{Powder Diffraction}} \bibinfo{volume}{29}, \bibinfo{number}{S2} (\bibinfo{year}{2014}), \bibinfo{pages}{S48--S64}.
\newblock


\bibitem[Banerjee and Lavie(2005)]%
        {banerjee-lavie-2005-meteor}
\bibfield{author}{\bibinfo{person}{Satanjeev Banerjee} {and} \bibinfo{person}{Alon Lavie}.} \bibinfo{year}{2005}\natexlab{}.
\newblock \showarticletitle{{METEOR}: An Automatic Metric for {MT} Evaluation with Improved Correlation with Human Judgments}. In \bibinfo{booktitle}{\emph{Proceedings of the {ACL} Workshop on Intrinsic and Extrinsic Evaluation Measures for Machine Translation and/or Summarization}}. \bibinfo{publisher}{Association for Computational Linguistics}, \bibinfo{address}{Ann Arbor, Michigan}, \bibinfo{pages}{65--72}.
\newblock
\urldef\tempurl%
\url{https://aclanthology.org/W05-0909}
\showURL{%
\tempurl}


\bibitem[Barone and Sennrich(2017)]%
        {barone2017parallel}
\bibfield{author}{\bibinfo{person}{Antonio Valerio~Miceli Barone} {and} \bibinfo{person}{Rico Sennrich}.} \bibinfo{year}{2017}\natexlab{}.
\newblock \showarticletitle{A parallel corpus of python functions and documentation strings for automated code documentation and code generation}.
\newblock \bibinfo{journal}{\emph{arXiv preprint arXiv:1707.02275}} (\bibinfo{year}{2017}).
\newblock


\bibitem[Beyene(2021)]%
        {beyene2021source}
\bibfield{author}{\bibinfo{person}{Michael Beyene}.} \bibinfo{year}{2021}\natexlab{}.
\newblock \emph{\bibinfo{title}{Source-code Summarization of Java Methods Using Control-Flow Graphs}}.
\newblock \bibinfo{thesistype}{Ph.\,D. Dissertation}. \bibinfo{school}{University of Saskatchewan}.
\newblock


\bibitem[Biswas et~al\mbox{.}(2019)]%
        {boa}
\bibfield{author}{\bibinfo{person}{Sumon Biswas}, \bibinfo{person}{Md~Johirul Islam}, \bibinfo{person}{Yijia Huang}, {and} \bibinfo{person}{Hridesh Rajan}.} \bibinfo{year}{2019}\natexlab{}.
\newblock \showarticletitle{Boa Meets Python: A Boa Dataset of Data Science Software in Python Language}. In \bibinfo{booktitle}{\emph{MSR'19: 16th International Conference on Mining Software Repositories}} (Montreal, Canada).
\newblock


\bibitem[Browning et~al\mbox{.}(2014)]%
        {browning2014docstring}
\bibfield{author}{\bibinfo{person}{J~Burton Browning}, \bibinfo{person}{Marty Alchin}, \bibinfo{person}{J~Burton Browning}, {and} \bibinfo{person}{Marty Alchin}.} \bibinfo{year}{2014}\natexlab{}.
\newblock \showarticletitle{Docstring Conventions}.
\newblock \bibinfo{journal}{\emph{Pro Python: Second Edition}} (\bibinfo{year}{2014}), \bibinfo{pages}{335--340}.
\newblock


\bibitem[Chyung et~al\mbox{.}(2017)]%
        {chyung2017evidence}
\bibfield{author}{\bibinfo{person}{Seung~Youn Chyung}, \bibinfo{person}{Katherine Roberts}, \bibinfo{person}{Ieva Swanson}, {and} \bibinfo{person}{Andrea Hankinson}.} \bibinfo{year}{2017}\natexlab{}.
\newblock \showarticletitle{Evidence-based survey design: The use of a midpoint on the Likert scale}.
\newblock \bibinfo{journal}{\emph{Performance Improvement}} \bibinfo{volume}{56}, \bibinfo{number}{10} (\bibinfo{year}{2017}), \bibinfo{pages}{15--23}.
\newblock


\bibitem[Clement et~al\mbox{.}(2020)]%
        {clement2020pymt5}
\bibfield{author}{\bibinfo{person}{Colin~B Clement}, \bibinfo{person}{Dawn Drain}, \bibinfo{person}{Jonathan Timcheck}, \bibinfo{person}{Alexey Svyatkovskiy}, {and} \bibinfo{person}{Neel Sundaresan}.} \bibinfo{year}{2020}\natexlab{}.
\newblock \showarticletitle{PyMT5: multi-mode translation of natural language and Python code with transformers}.
\newblock \bibinfo{journal}{\emph{arXiv preprint arXiv:2010.03150}} (\bibinfo{year}{2020}).
\newblock


\bibitem[Cui et~al\mbox{.}(2022)]%
        {cui2022codeexp}
\bibfield{author}{\bibinfo{person}{Haotian Cui}, \bibinfo{person}{Chenglong Wang}, \bibinfo{person}{Junjie Huang}, \bibinfo{person}{Jeevana~Priya Inala}, \bibinfo{person}{Todd Mytkowicz}, \bibinfo{person}{Bo Wang}, \bibinfo{person}{Jianfeng Gao}, {and} \bibinfo{person}{Nan Duan}.} \bibinfo{year}{2022}\natexlab{}.
\newblock \showarticletitle{CodeExp: Explanatory Code Document Generation}.
\newblock \bibinfo{journal}{\emph{arXiv preprint arXiv:2211.15395}} (\bibinfo{year}{2022}).
\newblock


\bibitem[Feldt and Magazinius(2010)]%
        {feldt2010validity}
\bibfield{author}{\bibinfo{person}{Robert Feldt} {and} \bibinfo{person}{Ana Magazinius}.} \bibinfo{year}{2010}\natexlab{}.
\newblock \showarticletitle{Validity threats in empirical software engineering research-an initial survey.}. In \bibinfo{booktitle}{\emph{Seke}}. \bibinfo{pages}{374--379}.
\newblock


\bibitem[Feng et~al\mbox{.}(2020)]%
        {codebert}
\bibfield{author}{\bibinfo{person}{Zhangyin Feng}, \bibinfo{person}{Daya Guo}, \bibinfo{person}{Duyu Tang}, \bibinfo{person}{Nan Duan}, \bibinfo{person}{Xiaocheng Feng}, \bibinfo{person}{Ming Gong}, \bibinfo{person}{Linjun Shou}, \bibinfo{person}{Bing Qin}, \bibinfo{person}{Ting Liu}, \bibinfo{person}{Daxin Jiang}, {and} \bibinfo{person}{Ming Zhou}.} \bibinfo{year}{2020}\natexlab{}.
\newblock \showarticletitle{CodeBERT: {A} Pre-Trained Model for Programming and Natural Languages}.
\newblock \bibinfo{journal}{\emph{CoRR}}  \bibinfo{volume}{abs/2002.08155} (\bibinfo{year}{2020}).
\newblock
\showeprint[arXiv]{2002.08155}
\urldef\tempurl%
\url{https://arxiv.org/abs/2002.08155}
\showURL{%
\tempurl}


\bibitem[Grotov et~al\mbox{.}(2022)]%
        {grotov2022large}
\bibfield{author}{\bibinfo{person}{Konstantin Grotov}, \bibinfo{person}{Sergey Titov}, \bibinfo{person}{Vladimir Sotnikov}, \bibinfo{person}{Yaroslav Golubev}, {and} \bibinfo{person}{Timofey Bryksin}.} \bibinfo{year}{2022}\natexlab{}.
\newblock \showarticletitle{A large-scale comparison of Python code in Jupyter notebooks and scripts}. In \bibinfo{booktitle}{\emph{Proceedings of the 19th International Conference on Mining Software Repositories}}. \bibinfo{pages}{353--364}.
\newblock


\bibitem[Guo et~al\mbox{.}(2022)]%
        {unixcoder}
\bibfield{author}{\bibinfo{person}{Daya Guo}, \bibinfo{person}{Shuai Lu}, \bibinfo{person}{Nan Duan}, \bibinfo{person}{Yanlin Wang}, \bibinfo{person}{Ming Zhou}, {and} \bibinfo{person}{Jian Yin}.} \bibinfo{year}{2022}\natexlab{}.
\newblock \showarticletitle{UniXcoder: Unified Cross-Modal Pre-training for Code Representation}.
\newblock \bibinfo{journal}{\emph{arXiv preprint arXiv:2203.03850}} (\bibinfo{year}{2022}).
\newblock


\bibitem[Guo et~al\mbox{.}(2021)]%
        {guo2021graphcodebert}
\bibfield{author}{\bibinfo{person}{Daya Guo}, \bibinfo{person}{Shuo Ren}, \bibinfo{person}{Shuai Lu}, \bibinfo{person}{Zhangyin Feng}, \bibinfo{person}{Duyu Tang}, \bibinfo{person}{Shujie Liu}, \bibinfo{person}{Long Zhou}, \bibinfo{person}{Nan Duan}, \bibinfo{person}{Alexey Svyatkovskiy}, \bibinfo{person}{Shengyu Fu}, \bibinfo{person}{Michele Tufano}, \bibinfo{person}{Shao~Kun Deng}, \bibinfo{person}{Colin Clement}, \bibinfo{person}{Dawn Drain}, \bibinfo{person}{Neel Sundaresan}, \bibinfo{person}{Jian Yin}, \bibinfo{person}{Daxin Jiang}, {and} \bibinfo{person}{Ming Zhou}.} \bibinfo{year}{2021}\natexlab{}.
\newblock \bibinfo{title}{GraphCodeBERT: Pre-training Code Representations with Data Flow}.
\newblock
\newblock
\showeprint[arxiv]{2009.08366}~[cs.SE]


\bibitem[Haiduc et~al\mbox{.}(2010)]%
        {haiduc2010use}
\bibfield{author}{\bibinfo{person}{Sonia Haiduc}, \bibinfo{person}{Jairo Aponte}, \bibinfo{person}{Laura Moreno}, {and} \bibinfo{person}{Andrian Marcus}.} \bibinfo{year}{2010}\natexlab{}.
\newblock \showarticletitle{On the use of automated text summarization techniques for summarizing source code}. In \bibinfo{booktitle}{\emph{2010 17th Working conference on reverse engineering}}. IEEE, \bibinfo{pages}{35--44}.
\newblock


\bibitem[Hu et~al\mbox{.}(2018)]%
        {hu2018deep}
\bibfield{author}{\bibinfo{person}{Xing Hu}, \bibinfo{person}{Ge Li}, \bibinfo{person}{Xin Xia}, \bibinfo{person}{David Lo}, {and} \bibinfo{person}{Zhi Jin}.} \bibinfo{year}{2018}\natexlab{}.
\newblock \showarticletitle{Deep code comment generation}. In \bibinfo{booktitle}{\emph{Proceedings of the 26th conference on program comprehension}}. \bibinfo{pages}{200--210}.
\newblock


\bibitem[Iyer et~al\mbox{.}(2016)]%
        {iyer-etal-2016-summarizing}
\bibfield{author}{\bibinfo{person}{Srinivasan Iyer}, \bibinfo{person}{Ioannis Konstas}, \bibinfo{person}{Alvin Cheung}, {and} \bibinfo{person}{Luke Zettlemoyer}.} \bibinfo{year}{2016}\natexlab{}.
\newblock \showarticletitle{Summarizing Source Code using a Neural Attention Model}. In \bibinfo{booktitle}{\emph{Proceedings of the 54th Annual Meeting of the Association for Computational Linguistics (Volume 1: Long Papers)}}. \bibinfo{publisher}{Association for Computational Linguistics}, \bibinfo{address}{Berlin, Germany}, \bibinfo{pages}{2073--2083}.
\newblock
\urldef\tempurl%
\url{https://doi.org/10.18653/v1/P16-1195}
\showDOI{\tempurl}


\bibitem[Kalliamvakou et~al\mbox{.}(2014)]%
        {kalliamvakou2014promises}
\bibfield{author}{\bibinfo{person}{Eirini Kalliamvakou}, \bibinfo{person}{Georgios Gousios}, \bibinfo{person}{Kelly Blincoe}, \bibinfo{person}{Leif Singer}, \bibinfo{person}{Daniel~M German}, {and} \bibinfo{person}{Daniela Damian}.} \bibinfo{year}{2014}\natexlab{}.
\newblock \showarticletitle{The promises and perils of mining github}. In \bibinfo{booktitle}{\emph{Proceedings of the 11th working conference on mining software repositories}}. \bibinfo{pages}{92--101}.
\newblock


\bibitem[LeClair et~al\mbox{.}(2019)]%
        {leclair2019neural}
\bibfield{author}{\bibinfo{person}{Alexander LeClair}, \bibinfo{person}{Siyuan Jiang}, {and} \bibinfo{person}{Collin McMillan}.} \bibinfo{year}{2019}\natexlab{}.
\newblock \bibinfo{title}{A Neural Model for Generating Natural Language Summaries of Program Subroutines}.
\newblock
\newblock
\showeprint[arxiv]{1902.01954}~[cs.SE]


\bibitem[Lethbridge et~al\mbox{.}(2003)]%
        {lethbridge2003software}
\bibfield{author}{\bibinfo{person}{Timothy~C Lethbridge}, \bibinfo{person}{Janice Singer}, {and} \bibinfo{person}{Andrew Forward}.} \bibinfo{year}{2003}\natexlab{}.
\newblock \showarticletitle{How software engineers use documentation: The state of the practice}.
\newblock \bibinfo{journal}{\emph{IEEE software}} \bibinfo{volume}{20}, \bibinfo{number}{6} (\bibinfo{year}{2003}), \bibinfo{pages}{35--39}.
\newblock


\bibitem[Luo et~al\mbox{.}(2018)]%
        {luo2018recognizing}
\bibfield{author}{\bibinfo{person}{Yang Luo}, \bibinfo{person}{Wanwangying Ma}, \bibinfo{person}{Yanhui Li}, \bibinfo{person}{Zhifei Chen}, {and} \bibinfo{person}{Lin Chen}.} \bibinfo{year}{2018}\natexlab{}.
\newblock \showarticletitle{Recognizing potential runtime types from python docstrings}. In \bibinfo{booktitle}{\emph{Software Analysis, Testing, and Evolution: 8th International Conference, SATE 2018, Shenzhen, Guangdong, China, November 23--24, 2018, Proceedings 8}}. Springer, \bibinfo{pages}{68--84}.
\newblock


\bibitem[McBurney(2015)]%
        {mcburney2015automatic}
\bibfield{author}{\bibinfo{person}{Paul~W McBurney}.} \bibinfo{year}{2015}\natexlab{}.
\newblock \showarticletitle{Automatic documentation generation via source code summarization}. In \bibinfo{booktitle}{\emph{2015 IEEE/ACM 37th IEEE International Conference on Software Engineering}}, Vol.~\bibinfo{volume}{2}. IEEE, \bibinfo{pages}{903--906}.
\newblock


\bibitem[McBurney and McMillan(2014)]%
        {mcburney2014automatic}
\bibfield{author}{\bibinfo{person}{Paul~W McBurney} {and} \bibinfo{person}{Collin McMillan}.} \bibinfo{year}{2014}\natexlab{}.
\newblock \showarticletitle{Automatic documentation generation via source code summarization of method context}. In \bibinfo{booktitle}{\emph{Proceedings of the 22nd International Conference on Program Comprehension}}. \bibinfo{pages}{279--290}.
\newblock


\bibitem[McHugh(2012)]%
        {mchugh2012interrater}
\bibfield{author}{\bibinfo{person}{Mary~L McHugh}.} \bibinfo{year}{2012}\natexlab{}.
\newblock \showarticletitle{Interrater reliability: the kappa statistic}.
\newblock \bibinfo{journal}{\emph{Biochemia medica}} \bibinfo{volume}{22}, \bibinfo{number}{3} (\bibinfo{year}{2012}), \bibinfo{pages}{276--282}.
\newblock


\bibitem[Meng et~al\mbox{.}(2022)]%
        {meng2022generating}
\bibfield{author}{\bibinfo{person}{Yu Meng}, \bibinfo{person}{Jiaxin Huang}, \bibinfo{person}{Yu Zhang}, {and} \bibinfo{person}{Jiawei Han}.} \bibinfo{year}{2022}\natexlab{}.
\newblock \showarticletitle{Generating training data with language models: Towards zero-shot language understanding}.
\newblock \bibinfo{journal}{\emph{Advances in Neural Information Processing Systems}}  \bibinfo{volume}{35} (\bibinfo{year}{2022}), \bibinfo{pages}{462--477}.
\newblock


\bibitem[Mir et~al\mbox{.}(2022)]%
        {mir2022type4py}
\bibfield{author}{\bibinfo{person}{Amir~M Mir}, \bibinfo{person}{Evaldas Lato{\v{s}}kinas}, \bibinfo{person}{Sebastian Proksch}, {and} \bibinfo{person}{Georgios Gousios}.} \bibinfo{year}{2022}\natexlab{}.
\newblock \showarticletitle{Type4py: Practical deep similarity learning-based type inference for python}. In \bibinfo{booktitle}{\emph{Proceedings of the 44th International Conference on Software Engineering}}. \bibinfo{pages}{2241--2252}.
\newblock


\bibitem[Monperrus et~al\mbox{.}(2012)]%
        {monperrus2012should}
\bibfield{author}{\bibinfo{person}{Martin Monperrus}, \bibinfo{person}{Michael Eichberg}, \bibinfo{person}{Elif Tekes}, {and} \bibinfo{person}{Mira Mezini}.} \bibinfo{year}{2012}\natexlab{}.
\newblock \showarticletitle{What should developers be aware of? An empirical study on the directives of API documentation}.
\newblock \bibinfo{journal}{\emph{Empirical Software Engineering}}  \bibinfo{volume}{17} (\bibinfo{year}{2012}), \bibinfo{pages}{703--737}.
\newblock


\bibitem[Nguyen-Hoan et~al\mbox{.}(2010)]%
        {nguyen2010survey}
\bibfield{author}{\bibinfo{person}{Luke Nguyen-Hoan}, \bibinfo{person}{Shayne Flint}, {and} \bibinfo{person}{Ramesh Sankaranarayana}.} \bibinfo{year}{2010}\natexlab{}.
\newblock \showarticletitle{A survey of scientific software development}. In \bibinfo{booktitle}{\emph{Proceedings of the 2010 ACM-IEEE international symposium on empirical software engineering and measurement}}. \bibinfo{pages}{1--10}.
\newblock


\bibitem[Niu et~al\mbox{.}(2022a)]%
        {niu2022deep}
\bibfield{author}{\bibinfo{person}{Changan Niu}, \bibinfo{person}{Chuanyi Li}, \bibinfo{person}{Bin Luo}, {and} \bibinfo{person}{Vincent Ng}.} \bibinfo{year}{2022}\natexlab{a}.
\newblock \bibinfo{title}{Deep Learning Meets Software Engineering: A Survey on Pre-Trained Models of Source Code}.
\newblock
\newblock
\showeprint[arxiv]{2205.11739}~[cs.SE]


\bibitem[Niu et~al\mbox{.}(2023)]%
        {niu2023empirical}
\bibfield{author}{\bibinfo{person}{Changan Niu}, \bibinfo{person}{Chuanyi Li}, \bibinfo{person}{Vincent Ng}, \bibinfo{person}{Dongxiao Chen}, \bibinfo{person}{Jidong Ge}, {and} \bibinfo{person}{Bin Luo}.} \bibinfo{year}{2023}\natexlab{}.
\newblock \showarticletitle{An empirical comparison of pre-trained models of source code}.
\newblock \bibinfo{journal}{\emph{arXiv preprint arXiv:2302.04026}} (\bibinfo{year}{2023}).
\newblock


\bibitem[Niu et~al\mbox{.}(2022b)]%
        {niu2022sptcode}
\bibfield{author}{\bibinfo{person}{Changan Niu}, \bibinfo{person}{Chuanyi Li}, \bibinfo{person}{Vincent Ng}, \bibinfo{person}{Jidong Ge}, \bibinfo{person}{Liguo Huang}, {and} \bibinfo{person}{Bin Luo}.} \bibinfo{year}{2022}\natexlab{b}.
\newblock \bibinfo{title}{SPT-Code: Sequence-to-Sequence Pre-Training for Learning Source Code Representations}.
\newblock
\newblock
\showeprint[arxiv]{2201.01549}~[cs.SE]


\bibitem[Oliphant(2007)]%
        {oliphant2007python}
\bibfield{author}{\bibinfo{person}{Travis~E Oliphant}.} \bibinfo{year}{2007}\natexlab{}.
\newblock \showarticletitle{Python for scientific computing}.
\newblock \bibinfo{journal}{\emph{Computing in science \& engineering}} \bibinfo{volume}{9}, \bibinfo{number}{3} (\bibinfo{year}{2007}), \bibinfo{pages}{10--20}.
\newblock


\bibitem[OpenAI(2023)]%
        {OpenAI2023GPT4TR}
\bibfield{author}{\bibinfo{person}{OpenAI}.} \bibinfo{year}{2023}\natexlab{}.
\newblock \showarticletitle{GPT-4 Technical Report}.
\newblock \bibinfo{journal}{\emph{ArXiv}}  \bibinfo{volume}{abs/2303.08774} (\bibinfo{year}{2023}).
\newblock


\bibitem[Papineni et~al\mbox{.}(2002)]%
        {papineni-etal-2002-bleu}
\bibfield{author}{\bibinfo{person}{Kishore Papineni}, \bibinfo{person}{Salim Roukos}, \bibinfo{person}{Todd Ward}, {and} \bibinfo{person}{Wei-Jing Zhu}.} \bibinfo{year}{2002}\natexlab{}.
\newblock \showarticletitle{{B}leu: a Method for Automatic Evaluation of Machine Translation}. In \bibinfo{booktitle}{\emph{Proceedings of the 40th Annual Meeting of the Association for Computational Linguistics}}. \bibinfo{publisher}{Association for Computational Linguistics}, \bibinfo{address}{Philadelphia, Pennsylvania, USA}, \bibinfo{pages}{311--318}.
\newblock
\urldef\tempurl%
\url{https://doi.org/10.3115/1073083.1073135}
\showDOI{\tempurl}


\bibitem[Pawlik et~al\mbox{.}(2012)]%
        {pawlik2012documentation}
\bibfield{author}{\bibinfo{person}{Aleksandra Pawlik}, \bibinfo{person}{Judith Segal}, {and} \bibinfo{person}{Marian Petre}.} \bibinfo{year}{2012}\natexlab{}.
\newblock \showarticletitle{Documentation practices in scientific software development}. In \bibinfo{booktitle}{\emph{2012 5th International Workshop on Co-operative and Human Aspects of Software Engineering (CHASE)}}. IEEE, \bibinfo{pages}{113--119}.
\newblock


\bibitem[Peng et~al\mbox{.}(2022)]%
        {peng22hityper}
\bibfield{author}{\bibinfo{person}{Yun Peng}, \bibinfo{person}{Cuiyun Gao}, \bibinfo{person}{Zongjie Li}, \bibinfo{person}{Bowei Gao}, \bibinfo{person}{David Lo}, \bibinfo{person}{Qirun Zhang}, {and} \bibinfo{person}{Michael Lyu}.} \bibinfo{year}{2022}\natexlab{}.
\newblock \showarticletitle{Static Inference Meets Deep Learning: A Hybrid Type Inference Approach for Python}. In \bibinfo{booktitle}{\emph{Proceedings of the 44th International Conference on Software Engineering}} (Pittsburgh, Pennsylvania) \emph{(\bibinfo{series}{ICSE '22})}. \bibinfo{publisher}{Association for Computing Machinery}, \bibinfo{address}{New York, NY, USA}, \bibinfo{pages}{2019–2030}.
\newblock
\showISBNx{9781450392211}
\urldef\tempurl%
\url{https://doi.org/10.1145/3510003.3510038}
\showDOI{\tempurl}


\bibitem[Phan et~al\mbox{.}(2021)]%
        {phan2021cotext}
\bibfield{author}{\bibinfo{person}{Long Phan}, \bibinfo{person}{Hieu Tran}, \bibinfo{person}{Daniel Le}, \bibinfo{person}{Hieu Nguyen}, \bibinfo{person}{James Anibal}, \bibinfo{person}{Alec Peltekian}, {and} \bibinfo{person}{Yanfang Ye}.} \bibinfo{year}{2021}\natexlab{}.
\newblock \bibinfo{title}{CoTexT: Multi-task Learning with Code-Text Transformer}.
\newblock
\newblock
\showeprint[arxiv]{2105.08645}~[cs.AI]


\bibitem[Pinto et~al\mbox{.}(2018)]%
        {pinto2018scientists}
\bibfield{author}{\bibinfo{person}{Gustavo Pinto}, \bibinfo{person}{Igor Wiese}, {and} \bibinfo{person}{Luiz~Felipe Dias}.} \bibinfo{year}{2018}\natexlab{}.
\newblock \showarticletitle{How do scientists develop scientific software? An external replication}. In \bibinfo{booktitle}{\emph{2018 IEEE 25th international conference on software analysis, evolution and reengineering (SANER)}}. IEEE, \bibinfo{pages}{582--591}.
\newblock


\bibitem[Pradel et~al\mbox{.}(2020)]%
        {pradel2020typewriter}
\bibfield{author}{\bibinfo{person}{Michael Pradel}, \bibinfo{person}{Georgios Gousios}, \bibinfo{person}{Jason Liu}, {and} \bibinfo{person}{Satish Chandra}.} \bibinfo{year}{2020}\natexlab{}.
\newblock \showarticletitle{Typewriter: Neural type prediction with search-based validation}. In \bibinfo{booktitle}{\emph{Proceedings of the 28th ACM Joint Meeting on European Software Engineering Conference and Symposium on the Foundations of Software Engineering}}. \bibinfo{pages}{209--220}.
\newblock


\bibitem[Pressman(2005)]%
        {pressman2005software}
\bibfield{author}{\bibinfo{person}{Roger~S Pressman}.} \bibinfo{year}{2005}\natexlab{}.
\newblock \bibinfo{booktitle}{\emph{Software engineering: a practitioner's approach}}.
\newblock \bibinfo{publisher}{Palgrave macmillan}.
\newblock


\bibitem[Qi et~al\mbox{.}(2021)]%
        {qi2021prophetnetx}
\bibfield{author}{\bibinfo{person}{Weizhen Qi}, \bibinfo{person}{Yeyun Gong}, \bibinfo{person}{Yu Yan}, \bibinfo{person}{Can Xu}, \bibinfo{person}{Bolun Yao}, \bibinfo{person}{Bartuer Zhou}, \bibinfo{person}{Biao Cheng}, \bibinfo{person}{Daxin Jiang}, \bibinfo{person}{Jiusheng Chen}, \bibinfo{person}{Ruofei Zhang}, \bibinfo{person}{Houqiang Li}, {and} \bibinfo{person}{Nan Duan}.} \bibinfo{year}{2021}\natexlab{}.
\newblock \bibinfo{title}{ProphetNet-X: Large-Scale Pre-training Models for English, Chinese, Multi-lingual, Dialog, and Code Generation}.
\newblock
\newblock
\showeprint[arxiv]{2104.08006}~[cs.CL]


\bibitem[Raffel et~al\mbox{.}(2020)]%
        {t5}
\bibfield{author}{\bibinfo{person}{Colin Raffel}, \bibinfo{person}{Noam Shazeer}, \bibinfo{person}{Adam Roberts}, \bibinfo{person}{Katherine Lee}, \bibinfo{person}{Sharan Narang}, \bibinfo{person}{Michael Matena}, \bibinfo{person}{Yanqi Zhou}, \bibinfo{person}{Wei Li}, {and} \bibinfo{person}{Peter~J. Liu}.} \bibinfo{year}{2020}\natexlab{}.
\newblock \bibinfo{title}{Exploring the Limits of Transfer Learning with a Unified Text-to-Text Transformer}.
\newblock
\newblock
\showeprint[arxiv]{1910.10683}~[cs.LG]


\bibitem[Raffel et~al\mbox{.}(2023)]%
        {raffel2023exploring}
\bibfield{author}{\bibinfo{person}{Colin Raffel}, \bibinfo{person}{Noam Shazeer}, \bibinfo{person}{Adam Roberts}, \bibinfo{person}{Katherine Lee}, \bibinfo{person}{Sharan Narang}, \bibinfo{person}{Michael Matena}, \bibinfo{person}{Yanqi Zhou}, \bibinfo{person}{Wei Li}, {and} \bibinfo{person}{Peter~J. Liu}.} \bibinfo{year}{2023}\natexlab{}.
\newblock \bibinfo{title}{Exploring the Limits of Transfer Learning with a Unified Text-to-Text Transformer}.
\newblock
\newblock
\showeprint[arxiv]{1910.10683}~[cs.LG]


\bibitem[Rani et~al\mbox{.}(2021a)]%
        {rani2021comments}
\bibfield{author}{\bibinfo{person}{Pooja Rani}, \bibinfo{person}{Suada Abukar}, \bibinfo{person}{Nataliia Stulova}, \bibinfo{person}{Alexandre Bergel}, {and} \bibinfo{person}{Oscar Nierstrasz}.} \bibinfo{year}{2021}\natexlab{a}.
\newblock \showarticletitle{Do comments follow commenting conventions? a case study in java and python}. In \bibinfo{booktitle}{\emph{2021 IEEE 21st International Working Conference on Source Code Analysis and Manipulation (SCAM)}}. IEEE, \bibinfo{pages}{165--169}.
\newblock


\bibitem[Rani et~al\mbox{.}(2021b)]%
        {rani2021developers}
\bibfield{author}{\bibinfo{person}{Pooja Rani}, \bibinfo{person}{Mathias Birrer}, \bibinfo{person}{Sebastiano Panichella}, \bibinfo{person}{Mohammad Ghafari}, {and} \bibinfo{person}{Oscar Nierstrasz}.} \bibinfo{year}{2021}\natexlab{b}.
\newblock \showarticletitle{What do developers discuss about code comments?}. In \bibinfo{booktitle}{\emph{2021 IEEE 21st International Working Conference on Source Code Analysis and Manipulation (SCAM)}}. IEEE, \bibinfo{pages}{153--164}.
\newblock


\bibitem[Segal(2007)]%
        {segal2007some}
\bibfield{author}{\bibinfo{person}{Judith Segal}.} \bibinfo{year}{2007}\natexlab{}.
\newblock \showarticletitle{Some problems of professional end user developers}. In \bibinfo{booktitle}{\emph{IEEE Symposium on Visual Languages and Human-Centric Computing (VL/HCC 2007)}}. IEEE, \bibinfo{pages}{111--118}.
\newblock


\bibitem[Shmerlin et~al\mbox{.}(2015)]%
        {shmerlin2015document}
\bibfield{author}{\bibinfo{person}{Yulia Shmerlin}, \bibinfo{person}{Irit Hadar}, \bibinfo{person}{Doron Kliger}, {and} \bibinfo{person}{Hayim Makabee}.} \bibinfo{year}{2015}\natexlab{}.
\newblock \showarticletitle{To document or not to document? An exploratory study on developers’ motivation to document code}. In \bibinfo{booktitle}{\emph{Advanced Information Systems Engineering Workshops: CAiSE 2015 International Workshops, Stockholm, Sweden, June 8-9, 2015, Proceedings 27}}. Springer, \bibinfo{pages}{100--106}.
\newblock


\bibitem[Sokolova and Lapalme(2009)]%
        {sokolova2009systematic}
\bibfield{author}{\bibinfo{person}{Marina Sokolova} {and} \bibinfo{person}{Guy Lapalme}.} \bibinfo{year}{2009}\natexlab{}.
\newblock \showarticletitle{A systematic analysis of performance measures for classification tasks}.
\newblock \bibinfo{journal}{\emph{Information processing \& management}} \bibinfo{volume}{45}, \bibinfo{number}{4} (\bibinfo{year}{2009}), \bibinfo{pages}{427--437}.
\newblock


\bibitem[Sridhara et~al\mbox{.}(2011)]%
        {sridhara2011generating}
\bibfield{author}{\bibinfo{person}{Giriprasad Sridhara}, \bibinfo{person}{Lori Pollock}, {and} \bibinfo{person}{K Vijay-Shanker}.} \bibinfo{year}{2011}\natexlab{}.
\newblock \showarticletitle{Generating parameter comments and integrating with method summaries}. In \bibinfo{booktitle}{\emph{2011 IEEE 19th International Conference on Program Comprehension}}. IEEE, \bibinfo{pages}{71--80}.
\newblock


\bibitem[Sul{\'\i}r and Porub{\"a}n(2017)]%
        {sulir2017generating}
\bibfield{author}{\bibinfo{person}{Mat{\'u}{\v{s}} Sul{\'\i}r} {and} \bibinfo{person}{Jaroslav Porub{\"a}n}.} \bibinfo{year}{2017}\natexlab{}.
\newblock \showarticletitle{Generating method documentation using concrete values from executions}. In \bibinfo{booktitle}{\emph{6th Symposium on Languages, Applications and Technologies (SLATE 2017)}}. Schloss Dagstuhl-Leibniz-Zentrum fuer Informatik.
\newblock


\bibitem[Usman et~al\mbox{.}(2017)]%
        {usman2017taxonomy}
\bibfield{author}{\bibinfo{person}{Muhammad Usman}, \bibinfo{person}{Ricardo Britto}, \bibinfo{person}{Jürgen Börstler}, {and} \bibinfo{person}{Emilia Mendes}.} \bibinfo{year}{2017}\natexlab{}.
\newblock \showarticletitle{Taxonomies in software engineering: A Systematic mapping study and a revised taxonomy development method}.
\newblock \bibinfo{journal}{\emph{Information and Software Technology}}  \bibinfo{volume}{85} (\bibinfo{year}{2017}), \bibinfo{pages}{43--59}.
\newblock
\showISSN{0950-5849}
\urldef\tempurl%
\url{https://doi.org/10.1016/j.infsof.2017.01.006}
\showDOI{\tempurl}


\bibitem[Vidoni(2022)]%
        {vidoni2022understanding}
\bibfield{author}{\bibinfo{person}{Melina Vidoni}.} \bibinfo{year}{2022}\natexlab{}.
\newblock \showarticletitle{Understanding roxygen package documentation in R}.
\newblock \bibinfo{journal}{\emph{Journal of Systems and Software}}  \bibinfo{volume}{188} (\bibinfo{year}{2022}), \bibinfo{pages}{111265}.
\newblock


\bibitem[Vidoni and Codabux(2023)]%
        {Vidoni2023taxonomy}
\bibfield{author}{\bibinfo{person}{Melina Vidoni} {and} \bibinfo{person}{Zadia Codabux}.} \bibinfo{year}{2023}\natexlab{}.
\newblock \showarticletitle{{Towards a Taxonomy of Roxygen Documentation in R Packages}}.
\newblock \bibinfo{journal}{\emph{Empirical Software Engineering}}  \bibinfo{volume}{XX} (\bibinfo{year}{2023}), \bibinfo{pages}{XX}.
\newblock
\showISSN{1573-7616}


\bibitem[Wang et~al\mbox{.}(2022)]%
        {wang2022pre}
\bibfield{author}{\bibinfo{person}{Haifeng Wang}, \bibinfo{person}{Jiwei Li}, \bibinfo{person}{Hua Wu}, \bibinfo{person}{Eduard Hovy}, {and} \bibinfo{person}{Yu Sun}.} \bibinfo{year}{2022}\natexlab{}.
\newblock \showarticletitle{Pre-trained language models and their applications}.
\newblock \bibinfo{journal}{\emph{Engineering}} (\bibinfo{year}{2022}).
\newblock


\bibitem[Wang et~al\mbox{.}(2021)]%
        {codet5}
\bibfield{author}{\bibinfo{person}{Yue Wang}, \bibinfo{person}{Weishi Wang}, \bibinfo{person}{Shafiq~R. Joty}, {and} \bibinfo{person}{Steven C.~H. Hoi}.} \bibinfo{year}{2021}\natexlab{}.
\newblock \showarticletitle{CodeT5: Identifier-aware Unified Pre-trained Encoder-Decoder Models for Code Understanding and Generation}.
\newblock \bibinfo{journal}{\emph{CoRR}}  \bibinfo{volume}{abs/2109.00859} (\bibinfo{year}{2021}).
\newblock
\showeprint[arXiv]{2109.00859}
\urldef\tempurl%
\url{https://arxiv.org/abs/2109.00859}
\showURL{%
\tempurl}


\bibitem[Woo et~al\mbox{.}(2018)]%
        {woo2018cbam}
\bibfield{author}{\bibinfo{person}{Sanghyun Woo}, \bibinfo{person}{Jongchan Park}, \bibinfo{person}{Joon-Young Lee}, {and} \bibinfo{person}{In~So Kweon}.} \bibinfo{year}{2018}\natexlab{}.
\newblock \bibinfo{title}{CBAM: Convolutional Block Attention Module}.
\newblock
\newblock
\showeprint[arxiv]{1807.06521}~[cs.CV]


\bibitem[Zhang et~al\mbox{.}(2022)]%
        {zhang2022code}
\bibfield{author}{\bibinfo{person}{Haiyin Zhang}, \bibinfo{person}{Lu{\'\i}s Cruz}, {and} \bibinfo{person}{Arie Van~Deursen}.} \bibinfo{year}{2022}\natexlab{}.
\newblock \showarticletitle{Code smells for machine learning applications}. In \bibinfo{booktitle}{\emph{Proceedings of the 1st International Conference on AI Engineering: Software Engineering for AI}}. \bibinfo{pages}{217--228}.
\newblock


\bibitem[Zheng et~al\mbox{.}(2019)]%
        {zheng2019codeattention}
\bibfield{author}{\bibinfo{person}{Wenhao Zheng}, \bibinfo{person}{Hongyu Zhou}, \bibinfo{person}{Ming Li}, {and} \bibinfo{person}{Jianxin Wu}.} \bibinfo{year}{2019}\natexlab{}.
\newblock \showarticletitle{CodeAttention: translating source code to comments by exploiting the code constructs}.
\newblock \bibinfo{journal}{\emph{Frontiers of Computer Science}}  \bibinfo{volume}{13} (\bibinfo{year}{2019}), \bibinfo{pages}{565--578}.
\newblock


\end{thebibliography}

\end{document}